\documentclass[letterpaper]{article}

\usepackage[T1]{fontenc}
\usepackage{enumitem}

\usepackage{geometry}
\usepackage{setspace}

\usepackage[style = chem-acs]{biblatex}
\usepackage{graphicx}
\usepackage{float}
\newfloat{scheme}{htbp}{los}
\floatname{scheme}{Scheme}
\floatname{chart}{Chart}
\newfloat{graph}{htbp}{loh}

\usepackage{chemformula} 
\usepackage[version = 4]{mhchem} 

\usepackage{authblk}
\author[1]{Eduard A. Podshivaylov\textsuperscript{\textdagger}}
\author[1]{Egor B. Kurgin\textsuperscript{\textdagger}}
\affil[1]{Voevodsky Institute of Chemical Kinetics and Combustion SB RAS, Novosibirsk 630090, Russia}
\author[2]{Changgang Yang}
\author[2]{Guofeng Zhang}
\affil[2]{State Key Laboratory of Quantum Optics and Quantum Optics Devices, Institute of Laser Spectroscopy, Collaborative Innovation Center of Extreme Optics, Shanxi University, Taiyuan, 030006 China}
\author[1]{Pavel A. Frantsuzov*}

\title{Quantifying Blinking Suppression and Probing Its Mechanism in PPD-Treated Single Quantum Dots}

\date{*Email: frantsuzov@kinetics.nsc.ru \\ \textsuperscript{\textdagger}These authors contributed equally to this work.}

\begin{document}

\maketitle
\newpage

\begin{abstract}
Colloidal semiconductor quantum dots (QDs) are highly promising materials for various applications, yet photoluminescence intermittency (blinking) remains a key obstacle to their practical use. While recent advances have enabled partial or complete suppression of this effect, both the degree of suppression and the mechanism by which it occurs are often difficult to establish. In the present work we analyse the blinking suppression of single CdSe/ZnS QDs after treatment with p-phenylenediamine (PPD), recently discovered by Yang et al. We revisit the blinking mechanism in this system and show that it is governed predominantly by the trapping mechanism, rather than by Auger ionization as previously proposed. For data analysis we apply a threshold-free two-dimensional criterion recently proposed by Podshivaylov and Frantsuzov. The criterion is based on the relative quantum yield estimator $\langle \hat{Q} \rangle^{(s)}$ and the decay-rate fluctuation parameter $\hat{\xi}$. We found that PPD treatment significantly increases the relative quantum yield and leads to only a modest increase in $\hat{\xi}$, indicating that the dominant effect of PPD is to shift the distribution of the non-radiative recombination rate toward lower values, rather than to eliminate its fluctuations. We further suggest a number of possible mechanisms of PPD-induced suppression and propose critical experiments for their verification.
\end{abstract}

\newpage
Colloidal  semiconductor quantum dots (QDs) and nanocrystals (NCs) are versatile nanomaterials with applications spanning optoelectronics, quantum technologies, and biomedical imaging (see, \textit{e.g.}, reviews in Refs.~\cite{Agarwal2023,Rempel2024,Kim2025}). However, their practical performance as single-photon sources, fluorescent markers, and other single-emitter devices is limited by photoluminescence intermittency~\autocite{gee2025} - commonly known as blinking~\autocite{FrantsuzovNaturePhys2008,FrantsuzovNL2013,Efros2016,Yuan2018,Rabouw2019,BusovOS2019,Yang2025} - along with related instabilities such as spectral diffusion~\autocite{Gao2019,Podshivaylov2019,Hinterding2021,Mangnus2023,Conradt2023} and photodegradation~\autocite{An2018,baitova2023,Liu2019,Wahl2022}. Considerable progress has been made in suppressing these effects through advanced synthesis and surface engineering (see reviews in Refs. ~\cite{Yang2025}, ~\cite{Ahmad2025} and references therein). Yet, a fundamental challenge remains: how does one quantify the degree of suppression when blinking is reduced but not eliminated, and how can the mechanism responsible for the suppression be identified?

The dominant blinking mechanism in colloidal QDs is the so-called trapping mechanism (TM), suggested by Frantsuzov and Marcus~\autocite{FrantsuzovPRB2005}, in which fluctuations of the nonradiative recombination rate arise from carrier trapping at surface defects. Other mechanisms, such as Auger blinking (A-type) and hot-carrier blinking, may also contribute~\autocite{Yang2025}. A key consequence of the TM mechanism is that the emission intensity distribution is continuous, which makes the threshold-based separation into ON and OFF states inherently ambiguous.

Traditionally, blinking analysis has relied on intensity thresholding to segment trajectories into ON and OFF states, followed by statistical analysis of dwell-time distributions. Because of the continuous intensity distribution inherent to the TM mechanism, the choice of threshold is arbitrary and strongly affects the extracted parameters~\autocite{Frantsuzov2009}. An alternative is to examine the power spectral density (PSD) of the intensity trace, whose shape changes from a $1/f$~\autocite{Podshivaylov2023,Podshivaylov2026,Frantsuzov2009,FrantsuzovNL2013} dependence to a flat spectrum upon complete blinking suppression~\autocite{Praneeth2024}. However, as we will demonstrate below, the PSD becomes uninformative when only partial suppression is achieved - precisely the scenario most commonly encountered in practice.

Recently, Podshivaylov and Frantsuzov proposed a new threshold-free criterion that addresses these limitations~\autocite{Podshivaylov2026_criterion}. The criterion maps each quantum dot onto a two-dimensional plane of the relative quantum yield estimator $\langle \hat{Q} \rangle^{(s)}$ and the normalized decay-rate fluctuation parameter $\hat{\xi}$. It provides both a quantitative measure of blinking suppression and a preliminary classification of the underlying blinking mechanism. The method was validated on extensive simulated data representing all major blinking mechanisms. In the present work, we apply this criterion to experimental data, with the aim of not only quantifying the degree of suppression but also identifying the mechanism by which it occurs.

We focus on single CdSe/ZnS core/shell quantum dots capped with 1-dodecanethiol, in the absence and presence of p-phenylenediamine (PPD) ligands.
As was found by Yang et al.~\autocite{yang2021}, PPD treatment reduces the duration of OFF states in this system. In Ref.~\cite{yang2021}  the blinking was attributed to Auger ionization of the nanocrystal, while its suppression was explained by the trapping of a hot trion hole onto the PPD energy level.
In this work we present extended analysis of the experimental data including the new criterion, in order to find the physical mechanism underlying the observed suppression. Our results suggest that an alternative mechanism may be more appropriate.

\begin{figure}[h!]
    \centering
    \includegraphics[width=.7\linewidth]{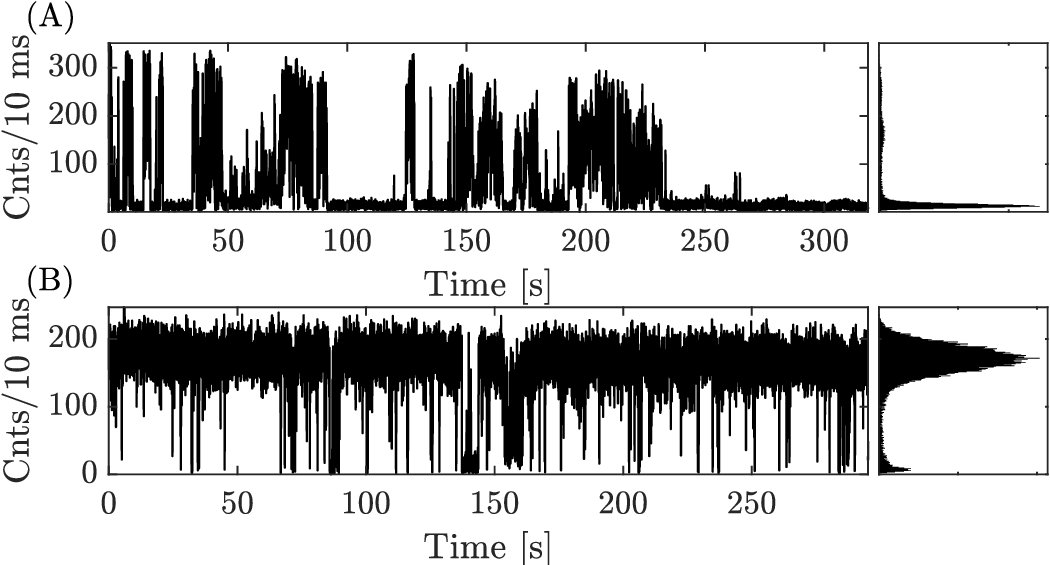}
    \caption{Typical blinking trajectories and corresponding photon distribution functions for single QDs (A) without and (B) with PPD.}
    \label{fig:trace_and_hist}
\end{figure}

For the present analysis, we selected only the set of quantum dots whose mean emission wavelength was 525 nm~\autocite{yang2021}; the other sets were not considered in this work. \textbf{Figure \ref{fig:trace_and_hist}} shows typical luminescence blinking trajectories and the corresponding photon distribution functions (PDFs) calculated from time-correlated single-photon counting (TCSPC) data for single CdSe/ZnS core/shell quantum dots without (A) and with (B) PPD. In the untreated case (A), the trajectory exhibits long OFF-state intervals, which manifest as a prominent peak near zero intensity in the PDF. In contrast, the presence of PPD (B) leads to a significant reduction in the overall OFF-state duration. This observed reduction constitutes the ``suppression'' of the blinking phenomenon.

The continuous shape of the photon distribution function indicates that blinking is governed by the trapping mechanism. The most established method to identify the specific blinking mechanism for single QDs and NCs is to construct a two-dimensional fluorescence lifetime--intensity distribution (FLID). The standard FLID method involves dividing a luminescence trajectory into several intensity levels and grouping the photons from each level into separate sets. The fluorescence lifetime for each set is then estimated by fitting the corresponding photoluminescence (PL) decay curve. Thus, each set is characterized by both its intensity, determined by the level, and an associated lifetime. In this work, we use a modified approach developed in Refs.~\cite{Podshivaylov2023} and~\cite{Podshivaylov2026}. Instead of using a monoexponential fit for the PL decay of each set, we employ a biexponential form to account for the delayed emission component observed in cadmium selenide-based QDs~\autocite{Rabouw2015,Hinterding2020}:

\begin{equation}
    \textrm{PL}(t) = C \left(\frac{A_\textrm{F}}{\tau_\textrm{F}} e^{-\frac{t}{\tau_\textrm{F}}} +\frac{A_\textrm{D}}{\tau_\textrm{D}} e^{-\frac{t}{\tau_\textrm{D}}}+ b\right)
    \label{Eq_FLID_PL}
\end{equation}
where $A_F$ and $\tau_F$ are the amplitude and decay time of the component contributing the larger fraction to the signal, $A_D$ and $\tau_D$ are the amplitude and decay time of the component contributing the smaller fraction, $b$ denotes the constant background level, and $C$ is a normalization constant from the maximum likelihood estimation procedure.

\begin{figure}[h!]
    \centering
    \includegraphics[width=.7\linewidth]{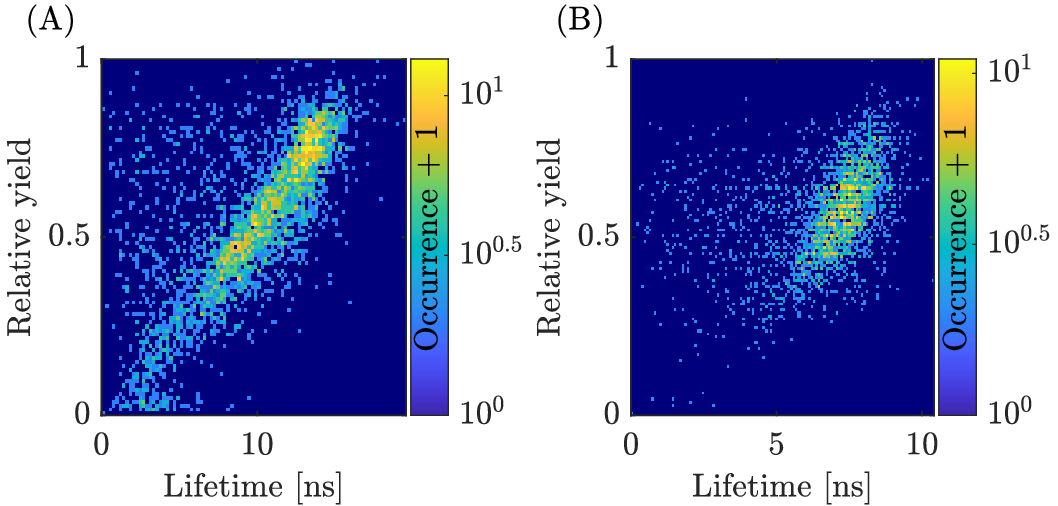}
    \caption{Fluorescence lifetime--intensity distributions for single QDs (A) without and (B) with PPD ligands.}
    \label{fig:FLID}
\end{figure}

Our analysis reveals that the overall contribution of the delayed emission component is almost negligible. However, the estimate for the fast decay time $\tau_F$ is more reliable. Therefore, for the construction of the FLID, we use the total intensity of each level and the corresponding fast decay time. The resulting FLIDs are presented in \textbf{Figure \ref{fig:FLID}} for QDs without (A) and with (B) PPD. In the untreated case (A), a clear, nearly linear correlation between the emission intensity and the fluorescence lifetime is observed. In the PPD-treated case (B), the FLID consists predominantly of a single distinct cluster corresponding to the ON state. However, this cluster retains a degree of linear dependence, as evidenced by its distinct slope. This behavior is consistent with the predictions of the trapping mechanism, which assumes a time-dependent nonradiative recombination rate, \textit{i.e.},
$$
k_\textrm{nr} = k_\textrm{nr}(t)
$$
where the time $t$ in this expression is assumed to be much greater than the characteristic fluorescence decay times. The corresponding time-dependent photoluminescence quantum yield (PLQY) is then given by:

\begin{equation}
    \textrm{PLQY} (t) = \frac{k_\textrm{r}}{k_\textrm{r} + k_\textrm{nr}(t)} \equiv k_\textrm{r} \tau(t)
    \label{PLQY}
\end{equation}
where $\tau$ is the luminescence lifetime and $k_r$ is the radiative recombination rate. Given the negligible contribution of the delayed component, we approximate the lifetime by the fast decay time, \textit{i.e.} $\tau \approx \tau_\textrm{F}$.

We note that in the earlier study ~\autocite{yang2021}, the blinking mechanism was inferred from the intensity--lifetime relation alone, without a full FLID analysis. The present FLID data, however, provide a more detailed picture and point toward the trapping mechanism. Other quantum dots from the same experiment, but with different emission wavelengths, exhibit different FLID patterns. In those cases, the FLID suggests a hybrid blinking mechanism, in which both TM and Auger processes contribute. A detailed investigation of these hybrid cases is beyond the scope of the present work and will be the subject of future studies.

\begin{figure}[h!]
		\centering
		\includegraphics[width=.7\linewidth]{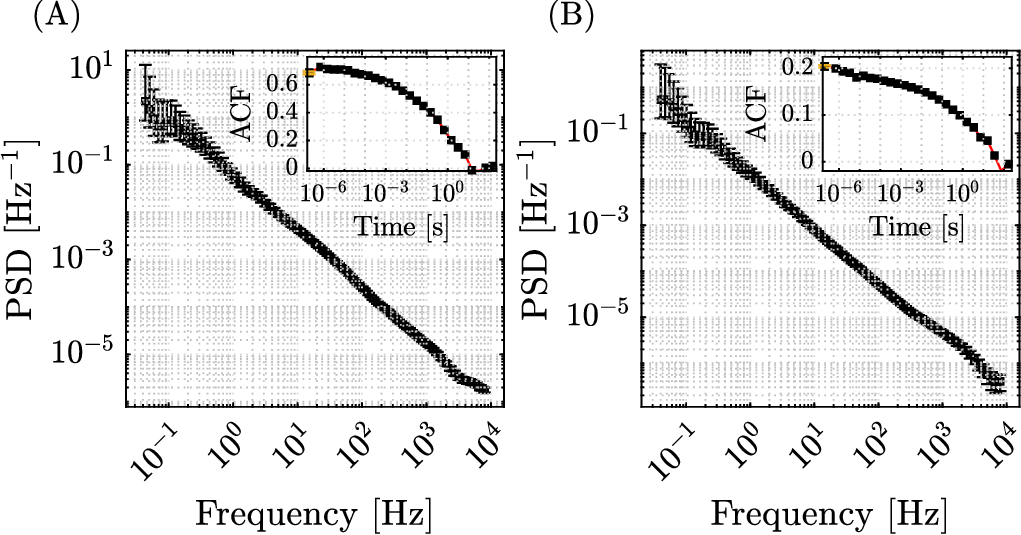}
		\caption{Power spectral density estimated using the modified Blackman-Tukey method for single QDs (A) without and (B) with PPD ligands. The error bars represent the 95\% confidence intervals. Insets show the corresponding autocorrelation functions (black squares) with their interpolations (red lines).}
		\label{fig:PSD}
\end{figure}

Another tool for analyzing blinking is the autocorrelation function (ACF) and its corresponding power spectral density. \textbf{Figure \ref{fig:PSD}} shows the PSD estimated using a modified Blackman--Tukey method recently proposed in Ref.~\cite{Podshivaylov2023}. The core idea of this method is to compute the Fourier transform of a windowed autocorrelation function, where the window itself is frequency-dependent. The ACFs, shown in the insets, were calculated following the procedure described in Ref.~\cite{Laurence2006}. The autocorrelation functions exhibit no characteristic decay times. Consistent with the trapping mechanism, the corresponding PSD also lacks a characteristic time scale, behaving as flicker noise with a spectral dependence close to $1/f$ for both the untreated QDs and the PPD-treated ones. Both cases exhibit the same qualitative dependence. This confirms the limitation anticipated in the Introduction: the PSD is uninformative when blinking is only partially suppressed, as in the case of PPD treatment where OFF times are reduced but not eliminated.

\begin{figure}[h!]
    \centering
    \includegraphics[width=.7\linewidth]{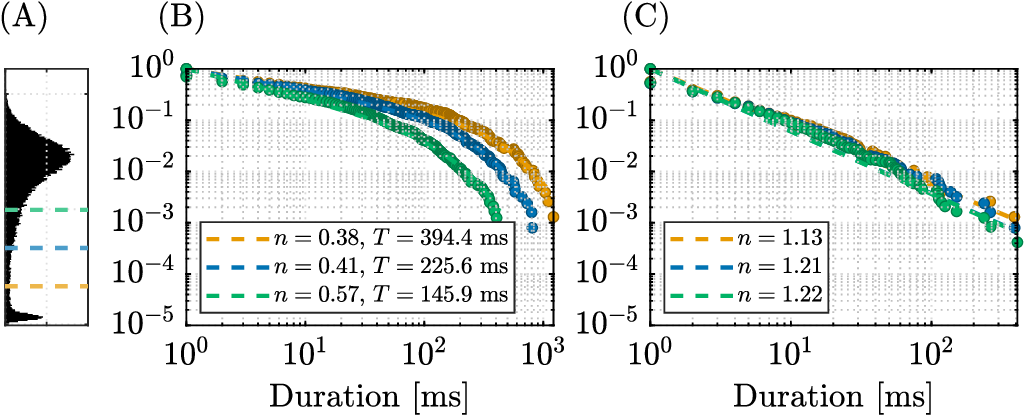}
    \caption{Threshold dependence of the ON/OFF dwell-time distributions. (A) Representative blinking trajectory with three different thresholds marked by colored lines. (B) Survival functions of the ON distributions for the three thresholds. (C) Survival functions of the OFF distributions for the three thresholds.}
    \label{fig:ON_OFF}
\end{figure}

To demonstrate how strongly the ON/OFF distributions depend on the chosen threshold, we analyzed them for three different threshold values: close to the zero-intensity peak, close to the ON-state peak, and midway between them. The dwell-time distributions were approximated by a power law with an exponential cutoff,
\begin{equation}
    P_\text{ON/OFF}(t) \propto t^{-(1+n)} e^{-t/T},
    \label{eqn:powerlaw}
\end{equation}
where $n$ is the power-law exponent and $T$ is the characteristic cutoff time. While the distribution is defined by Eq.~\eqref{eqn:powerlaw}, the fitting was performed on the corresponding survival functions to avoid binning artifacts. \textbf{Figure \ref{fig:ON_OFF}} shows a representative blinking trajectory with the three thresholds marked (panel A), along with the corresponding survival functions for the ON and OFF dwell times (panels B and C). It is clearly seen that the ON distribution changes its characteristic cutoff time $T$ significantly from threshold to threshold, whereas the OFF distribution shows only a modest change in the power-law exponent $n$. We note that the OFF distributions do not exhibit a detectable cutoff time within the measured range, so that the distribution follows a pure power law. This demonstrates that the choice of threshold introduces an unavoidable bias into the extracted blinking statistics.

To overcome the limitations of conventional methods demonstrated above, we apply the 2D criterion recently proposed in Ref.~\cite{Podshivaylov2026_criterion}. The method is based on the analysis of two quantities: the relative quantum yield estimator $\langle Q \rangle^{(s)}$ and the normalized decay-rate fluctuation parameter $\hat{\xi}$. In the notation of the original work, a hat denotes an estimated quantity and angle brackets denote the time average over the trajectory.

The parameter $\langle Q \rangle^{(s)}$ represents a self-consistent estimator of the relative quantum yield and is defined by the following relation:
\begin{equation}
    \langle Q \rangle^{(s)} = \frac{\langle\hat{n}\rangle}{\hat{N}_{\max}}
\end{equation}
where $\langle\hat{n}\rangle$ is the trajectory-averaged luminescence intensity, and $\hat{N}_{\max}$ is the estimator of the maximum luminescence intensity, given by
\begin{equation*}
    \hat{N}_{\max} = \frac{1}{4}\left( \sqrt{K^2 + 4\hat{n}_{\max}} - K \right)^2
\end{equation*}
where $\hat{n}_{\max}$ denotes the maximum intensity observed along the trajectory, and the parameter $K$ is defined as
\begin{equation*}
    K \approx \sqrt{2\ln M_{\max}}
\end{equation*}
with $M_{\max}$ being the number of bins attaining the maximum intensity, which can be practically estimated as $M_{\textrm{tot}}/10$ (see the original work for a detailed derivation).

\begin{figure}[h!]
    \centering
    \includegraphics[width=.7\linewidth]{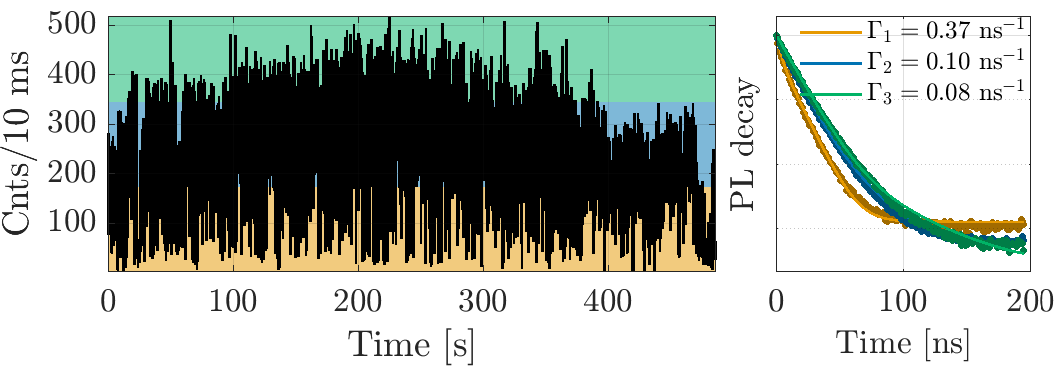}
    \caption{Blinking trajectory divided into three broad intensity levels (colored areas) and the corresponding fitted PL decay curves.}
    \label{fig:decay_fit_vis}
\end{figure}

To determine the parameter $\hat{\xi}$, one needs the distribution of the luminescence decay rates (the criterion can also be formulated in terms of the luminescence lifetimes $\tau = 1/\Gamma$). We obtain this distribution as follows. The intensity trajectory is divided into several equidistant levels, with a characteristic bin width of 2--3 photon counts, similar to the procedure used for the FLID construction. For each intensity level, a photoluminescence (PL) decay histogram is constructed, and the number of bins assigned to this level, $N_{\textrm{bins},i}$, is determined. This histogram is then fitted with a biexponential function using the maximum likelihood method, analogous to the expression given in \textbf{Equation \ref{Eq_FLID_PL}}:
\begin{equation*}
    \textrm{PL}(t) = C\left( A_\textrm{F}\Gamma_\textrm{F} e^{-\Gamma_\textrm{F} t} + A_\textrm{D}\Gamma_\textrm{D} e^{-\Gamma_\textrm{D} t} + b \right)
\end{equation*}
where $A_\textrm{F}$ and $\Gamma_\textrm{F} = 1/\tau_\textrm{F}$ are the amplitude and decay rate of the dominant component, $A_\textrm{D}$ and $\Gamma_\textrm{D} = 1/\tau_\textrm{D}$ are the amplitude and decay rate of the delayed component, $b$ denotes the constant background level, and $C$ is a normalization constant from the maximum likelihood estimation procedure. To illustrate the quality of the fit, \textbf{Figure \ref{fig:decay_fit_vis}} shows an example using three broad intensity levels instead of the numerous narrow levels used in the actual analysis. The luminescence decay rate is taken as the dominant component, $\Gamma = \Gamma_\textrm{F}$. To ensure proper normalization of the resulting distribution, each $\Gamma_i$ value is counted $N_{\textrm{bins},i}$ times.

The parameter $\hat{\xi}$ is then defined by the following expression:
\begin{equation}
    \hat{\xi} = 1 - \frac{\hat{\sigma}^2_\Gamma}{(\hat\Gamma_{\max} - \langle\hat\Gamma\rangle)(\langle\hat\Gamma\rangle - \hat\Gamma_{\min})}
\end{equation}
where $\hat\Gamma_{\max}$ and $\hat\Gamma_{\min}$ are the maximum and minimum values of $\Gamma$ over the distribution, $\langle\hat\Gamma\rangle$ is the mean value, and $\hat\sigma_\Gamma^2$ is the standard unbiased estimator of the variance.

\begin{figure}[h!]
    \centering
    \includegraphics[width=.7\linewidth]{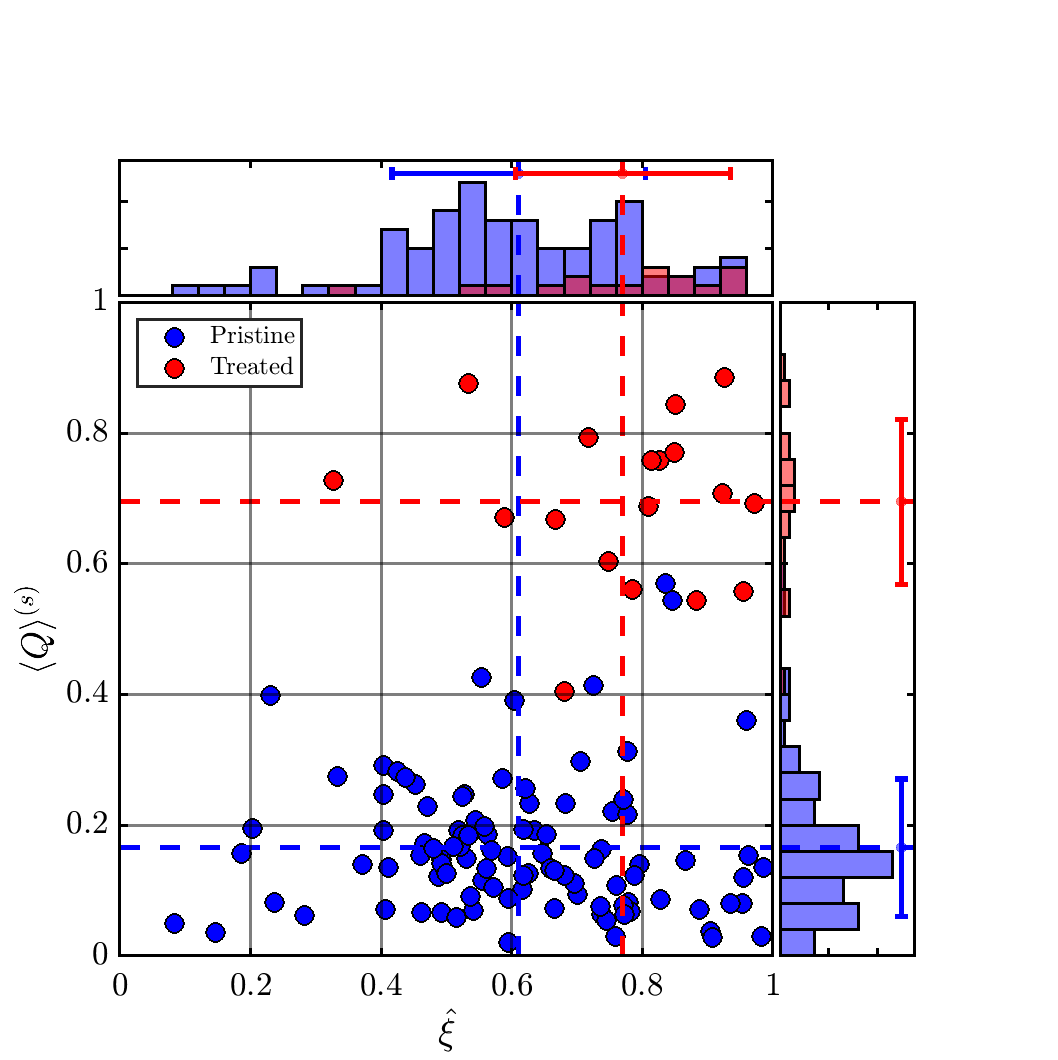}
    \caption{The $\hat{\xi}$ and $\langle Q \rangle^{(s)}$ values for each quantum dot from the set with PPD (red) and the set without PPD (blue).}
    \label{fig:criterion}
\end{figure}

\textbf{Figure \ref{fig:criterion}} presents the values of $\hat{\xi}$ and $\langle Q \rangle^{(s)}$ for two distinct sets of quantum dots: a control set without PPD and a set treated with PPD. Both sets of points lie in the central region of the $(\hat{\xi}, \langle Q \rangle^{(s)})$ plane, which is characteristic of the TM blinking mechanism or a hybrid mechanism, in agreement with the FLID analysis presented above.

The two-dimensional criterion allows us to address a fundamental question that cannot be answered quantitatively by conventional methods (such as FLID, PSD, or threshold analysis): does the PPD treatment primarily reduce the average level of non-radiative recombination, or does it mainly suppress the fluctuations of the non-radiative recombination rate? As seen in \textbf{Figure \ref{fig:criterion}}, the addition of PPD significantly increases the relative quantum yield $\langle Q \rangle^{(s)}$, while the decay-rate fluctuation parameter $\hat{\xi}$ changes only modestly. This directly indicates that the dominant effect of PPD is a shift in the distribution of the non-radiative recombination rate toward lower values, rather than the elimination of its fluctuations.

To understand this result in more detail, we analyze the distributions of the decay rate $\Gamma$. Indeed, for the decay rate, the following relation holds:
\begin{equation*}
    \Gamma = k_\textrm{r} + k_\textrm{nr}(t),
\end{equation*}
where $k_\textrm{r}$ is the radiative rate and $k_\textrm{nr}(t)$ is the fluctuating non-radiative rate. Consequently, the variance of $\Gamma$ is entirely determined by the fluctuations of $k_\textrm{nr}$:
\begin{equation*}
    \langle\langle\Gamma\rangle\rangle = \langle\langle k_\textrm{nr}(t) \rangle\rangle,
\end{equation*}
where the double angle brackets denote the cumulant (variance) of the corresponding quantity.

\begin{figure}[h!]
    \centering
    \includegraphics[width=.7\linewidth]{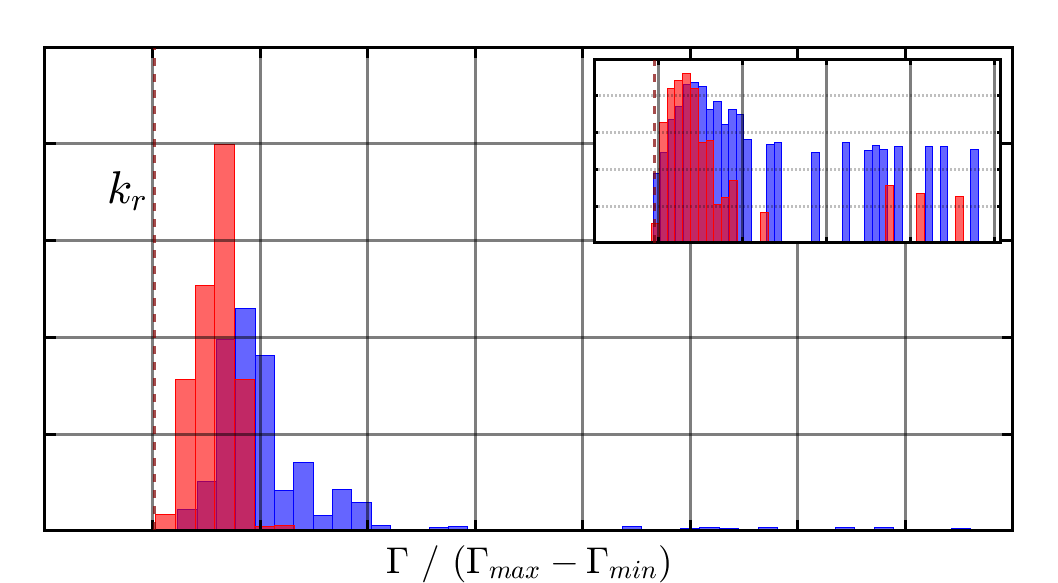}
    \caption{Distributions of the luminescence decay rate $\Gamma$ for representative QDs with PPD (red) and without PPD (blue). The horizontal axis is normalized to $\Gamma_{\max}-\Gamma_{\min}$. The dashed line marks the radiative rate $k_\textrm{r}$. The inset shows the same data on a logarithmic scale along the vertical axis.}
    \label{fig:Gamma_dist}
\end{figure}

\textbf{Figure \ref{fig:Gamma_dist}} shows the distributions of the decay rate $\Gamma$ for a representative QD without PPD (blue) and with PPD (red). The distributions are normalized to $\Gamma_{\max} - \Gamma_{\min}$, so that the horizontal axis is dimensionless. Two features are apparent. First, the main peak narrows slightly. Second, its center of mass shifts toward the radiative rate $k_\textrm{r}$, while a broad shoulder toward larger $\Gamma$ persists. This confirms that the mean value of \(k_\textrm{nr}\) decreases, whereas its variance does not change dramatically. This is fully consistent with the relatively modest change in $\hat{\xi}$ observed in \textbf{Figure \ref{fig:criterion}}.

Since the FLID analysis and the PSD behavior clearly indicate the TM mechanism, the explanation  based on Auger ionization and hot-hole trapping~\autocite{yang2021} does not apply to this system.
Therefore, new hypotheses are needed to explain the mechanism of blinking suppression due to PPD treatment. We consider several possible mechanisms below.

The recently proposed models of blinking within the trapping mechanism  were developed in Refs.~\cite{Podshivaylov2023} and~\cite{Podshivaylov2026} based on the electron-phonon coupling  variations (EPV).
These models successfully provide a quantitative description of the blinking properties for single semiconductor quantum dots~\autocite{Podshivaylov2023} and for single perovskite quantum dots~\autocite{Podshivaylov2026}.
In the models, the  nonradiative recombination is controlled by LO phonon assisted trapping of a charge carrier.
The trapping rate can be found using the Marcus--Jortner formula  (see for example Ref.~\cite{BarzykinACP02}).
In Ref.~\cite{Podshivaylov2023}, it was shown that for a deep trap, the trapping rate can be written as
\begin{equation}
    k_\textrm{nr} = k_0 S^{\alpha},
    \label{eqn:knr_model}
\end{equation}
where $k_0$ is a pre-exponential factor that depends on temperature, the coupling matrix element $V$,  and  the reorganization energy of acoustic modes. The exponent $\alpha$ is approximately equal to
the number of LO phonons that have to be emitted during the transition:
$$\alpha \approx \frac {\Delta E}{\hbar \omega}$$
where $\Delta E$ is the energy depth of the trap level and $\omega$ is the LO phonon frequency.
The Huang--Rhys parameter $S$ describes the strength of the electron--phonon coupling for LO phonons and is given by
\begin{equation}
    S = \frac{\lambda}{\hbar\omega},
    \label{eqn:S_def}
\end{equation}
where $\lambda$ is the reorganization energy of the LO phonons within the trapping. In the weak-coupling limit, relevant for CdSe, $S < 1$ (established for optical transitions, \textit{e.g.}, Ref.~\cite{lin2015}, whereas the value for defect trapping is not directly known).
Within the EPV models~\autocite{Podshivaylov2023,Podshivaylov2026}, the PL intensity blinking is a result of reorganization energy $\lambda(t)$ variations caused by the changes in the quantum dot crystal structure and
in the surrounding medium.   Several hypotheses for the PPD-induced suppression within EPV model can be suggested.\\

{\bf Hypothesis 1:} The suppression is a result of an increase the energy of the trap level $\Delta E$.
The trap energy increase can be result of two possible reasons:
\begin{enumerate}[label=\alph*)]
\item  The PPD molecule passivates the shallowest trap, so trapping occurs at a deeper one.
\item The ligand molecule generates local electric field which shifts the trap energy by the Stark effect.
\end{enumerate}

 This hypothesis could be tested by  absorption spectroscopy of the QD ensemble with and without PPD treatment.\\

{\bf Hypothesis 2:} The suppression is caused by a decrease in the pre-exponential factor  $k_0$.
This can be a result of two possible reasons:
\begin{enumerate}[label=\alph*)]
\item Decreasing of the coupling matrix element absolute value $|V|$ \\
\item Increasing of the acoustic modes reorganization energy.
\end{enumerate}

{\bf Hypothesis 3:} The PPD treatment causes a decrease in the reorganization energy  $\lambda(t)$.

 A change in $\lambda$ would be reflected in the phonon sidebands of the emission spectrum and in the Raman response of the nanocrystal, and could therefore be probed by phonon-sideband spectroscopy and resonant Raman measurements. \\

{\bf Hypothesis 4:} The PPD treatment decreases the rate of the competing process, namely  the trapping assisted by a vibrational mode of the ligand molecule.
 The energy of the C--H stretching modes (2800-3000 cm$^-1$) of the native ligand (1-dodecanethiol) matches the typical trap depth in CdSe nanocrystals.
 This makes the rate of the single-phonon assisted trapping process quite high.
As a result of adding PPD, its molecules can bind to defect sites.
Since the PPD molecule contains a significantly smaller number of C–H bonds, the trapping rate  via this pathway is substantially lower.
  The influence of ligand vibrational modes on the relaxation dynamics of excited states in single quantum dots has been demonstrated experimentally (see, \textit{e.g.}, Refs.~\cite{guyot2005,cooney2007,peterson2014}).

This hypothesis makes a directly testable prediction: replacement of 1-dodecanethiol with its deuterated analogue should shift the C--H mode to $\sim 2100$--$2200$~cm$^{-1}$ and break the resonance with the trap level. Because deuteration itself suppresses the vibrationally assisted trapping pathway, the deuterated samples are expected to exhibit significantly reduced blinking even before the addition of PPD. Consequently, the additional suppression produced by PPD on top of this effect should become much smaller or even negligible, as the relevant trapping pathway has already been weakened.

In this study, we have analyzed luminescence blinking in two sets of single CdSe/ZnS quantum dots---with and without PPD treatment. The presence of PPD leads to a significant reduction in both the number and duration of OFF states, yet standard analytical tools---threshold-based segmentation, FLID analysis, and PSD analysis---fail to provide a reliable quantitative measure of the degree of suppression. We have applied the recently proposed 2D criterion based on the estimators $\langle Q \rangle^{(s)}$ and $\hat{\xi}$ and have shown that it cleanly discriminates between the PPD-treated and untreated ensembles. Importantly, the criterion allows one to separate two aspects of suppression that are difficult to distinguish by conventional methods: a decrease in the average non-radiative recombination rate versus a reduction of its fluctuations. Our analysis shows that PPD treatment significantly increases the relative quantum yield \(\langle Q \rangle^{(s)}\), while the decay-rate fluctuation parameter \(\hat{\xi}\) changes only modestly. This indicates that the dominant effect of PPD is to shift the distribution of the non-radiative recombination rate toward lower values, while leaving its relative fluctuations largely unchanged.

Using the FLID data, we have revisited the mechanism of blinking in this system. The results indicate that the observed blinking is governed predominantly by the trapping mechanism, rather than by Auger ionization as previously proposed. On this basis, we have considered several possible mechanisms by which PPD suppresses blinking, and we have suggested four possible molecular mechanisms of the blinking suppression.

The proposed 2D criterion, combined with the additional data analysis, thus provides not only a quantitative measure of the degree of blinking suppression, but also a route toward identifying the physical mechanism responsible for it. The approach is simple, reproducible, and readily extendable to other quantum dot and nanocrystal systems where partial blinking suppression is achieved.

\section*{Acknowledgements}
E.B.K., E.A.P, and P.A.F.  acknowledge the core funding from the Russian Federal Ministry of Science and Higher Education (FWGF-2026-0007). C.Y. and G.Z. acknowledge the support from the Natural Science Foundation of China (Nos. 62127817, 62575162, 62505165) and the Natural Science Foundation of Shanxi Province (No. 202303021222031).

\printbibliography

\end{document}